\documentclass[twocolumn,preprintnumbers,superscriptaddress,showpacs,amsmath,amssymb]{revtex4}

\usepackage[dvips]{epsfig,color}
\include{graphics}

\usepackage{graphicx}
\usepackage{dcolumn}

\usepackage{rotating}

\begin{document}

\newcommand*{\GWU}{The George Washington University, Washington, DC 20052}
\affiliation{\GWU}
\newcommand*{\ANL}{Argonne National Laboratory, Argonne, Illinois 60441}
\newcommand*{\ANLindex}{1}
\affiliation{\ANL}
\newcommand*{\ASU}{Arizona State University, Tempe, Arizona 85287-1504}
\newcommand*{\ASUindex}{2}
\affiliation{\ASU}
\newcommand*{\UCLA}{University of California at Los Angeles, Los Angeles, California  90095-1547}
\newcommand*{\UCLAindex}{3}
\affiliation{\UCLA}
\newcommand*{\CSUDH}{California State University, Dominguez Hills, Carson, CA 90747}
\newcommand*{\CSUDHindex}{4}
\affiliation{\CSUDH}
\newcommand*{\CANISIUS}{Canisius College, Buffalo, NY}
\newcommand*{\CANISIUSindex}{5}
\affiliation{\CANISIUS}
\newcommand*{\CMU}{Carnegie Mellon University, Pittsburgh, Pennsylvania 15213}
\newcommand*{\CMUindex}{6}
\affiliation{\CMU}
\newcommand*{\CUA}{Catholic University of America, Washington, D.C. 20064}
\newcommand*{\CUAindex}{7}
\affiliation{\CUA}
\newcommand*{\SACLAY}{CEA, Centre de Saclay, Irfu/Service de Physique Nucl\'eaire, 91191 Gif-sur-Yvette, France}
\newcommand*{\SACLAYindex}{8}
\affiliation{\SACLAY}
\newcommand*{\CNU}{Christopher Newport University, Newport News, Virginia 23606}
\newcommand*{\CNUindex}{9}
\affiliation{\CNU}
\newcommand*{\UCONN}{University of Connecticut, Storrs, Connecticut 06269}
\newcommand*{\UCONNindex}{10}
\affiliation{\UCONN}
\newcommand*{\EDINBURGH}{Edinburgh University, Edinburgh EH9 3JZ, United Kingdom}
\newcommand*{\EDINBURGHindex}{11}
\affiliation{\EDINBURGH}
\newcommand*{\FU}{Fairfield University, Fairfield CT 06824}
\newcommand*{\FUindex}{12}
\affiliation{\FU}
\newcommand*{\FIU}{Florida International University, Miami, Florida 33199}
\newcommand*{\FIUindex}{13}
\affiliation{\FIU}
\newcommand*{\FSU}{Florida State University, Tallahassee, Florida 32306}
\newcommand*{\FSUindex}{14}
\affiliation{\FSU}
\newcommand*{\Genova}{Universit$\grave{a}$ di Genova, 16146 Genova, Italy}
\newcommand*{\Genovaindex}{15}
\affiliation{\Genova}
\newcommand*{\ISU}{Idaho State University, Pocatello, Idaho 83209}
\newcommand*{\ISUindex}{17}
\affiliation{\ISU}
\newcommand*{\INFNFR}{INFN, Laboratori Nazionali di Frascati, 00044 Frascati, Italy}
\newcommand*{\INFNFRindex}{18}
\affiliation{\INFNFR}
\newcommand*{\INFNGE}{INFN, Sezione di Genova, 16146 Genova, Italy}
\newcommand*{\INFNGEindex}{19}
\affiliation{\INFNGE}
\newcommand*{\INFNRO}{INFN, Sezione di Roma Tor Vergata, 00133 Rome, Italy}
\newcommand*{\INFNROindex}{20}
\affiliation{\INFNRO}
\newcommand*{\ORSAY}{Institut de Physique Nucl\'eaire ORSAY, Orsay, France}
\newcommand*{\ORSAYindex}{21}
\affiliation{\ORSAY}
\newcommand*{\ITEP}{Institute of Theoretical and Experimental Physics, Moscow, 117259, Russia}
\newcommand*{\ITEPindex}{22}
\affiliation{\ITEP}
\newcommand*{\JMU}{James Madison University, Harrisonburg, Virginia 22807}
\newcommand*{\JMUindex}{23}
\affiliation{\JMU}
\newcommand*{\KNU}{Kyungpook National University, Daegu 702-701, Republic of Korea}
\newcommand*{\KNUindex}{24}
\affiliation{\KNU}
\newcommand*{\LPSC}{LPSC, Universite Joseph Fourier, CNRS/IN2P3, INPG, Grenoble, France
}
\newcommand*{\LPSCindex}{25}
\affiliation{\LPSC}
\newcommand*{\MIT}{Massachusetts Institute of Technology, Cambridge, Massachusetts  02139-4307}
\newcommand*{\MITindex}{26}
\affiliation{\MIT}
\newcommand*{\UNH}{University of New Hampshire, Durham, New Hampshire 03824-3568}
\newcommand*{\UNHindex}{27}
\affiliation{\UNH}
\newcommand*{\NSU}{Norfolk State University, Norfolk, Virginia 23504}
\newcommand*{\NSUindex}{28}
\affiliation{\NSU}
\newcommand*{\OHIOU}{Ohio University, Athens, Ohio  45701}
\newcommand*{\OHIOUindex}{29}
\affiliation{\OHIOU}
\newcommand*{\ODU}{Old Dominion University, Norfolk, Virginia 23529}
\newcommand*{\ODUindex}{30}
\affiliation{\ODU}
\newcommand*{\RPI}{Rensselaer Polytechnic Institute, Troy, New York 12180-3590}
\newcommand*{\RPIindex}{31}
\affiliation{\RPI}
\newcommand*{\URICH}{University of Richmond, Richmond, Virginia 23173}
\newcommand*{\URICHindex}{32}
\affiliation{\URICH}
\newcommand*{\ROMAII}{Universita' di Roma Tor Vergata, 00133 Rome Italy}
\newcommand*{\ROMAIIindex}{33}
\affiliation{\ROMAII}
\newcommand*{\MSU}{Skobeltsyn Nuclear Physics Institute, Skobeltsyn Nuclear Physics Institute, 119899 Moscow, Russia}
\newcommand*{\MSUindex}{34}
\affiliation{\MSU}
\newcommand*{\SCAROLINA}{University of South Carolina, Columbia, South Carolina 29208}
\newcommand*{\SCAROLINAindex}{35}
\affiliation{\SCAROLINA}
\newcommand*{\JLAB}{Thomas Jefferson National Accelerator Facility, Newport News, Virginia 23606}
\newcommand*{\JLABindex}{36}
\affiliation{\JLAB}
\newcommand*{\UNIONC}{Union College, Schenectady, NY 12308}
\newcommand*{\UNIONCindex}{37}
\affiliation{\UNIONC}
\newcommand*{\UTFSM}{Universidad T\'{e}cnica Federico Santa Mar\'{i}a, Casilla 110-V Valpara\'{i}so, Chile}
\newcommand*{\UTFSMindex}{38}
\affiliation{\UTFSM}
\newcommand*{\GLASGOW}{University of Glasgow, Glasgow G12 8QQ, United Kingdom}
\newcommand*{\GLASGOWindex}{39}
\affiliation{\GLASGOW}
\newcommand*{\VIRGINIA}{University of Virginia, Charlottesville, Virginia 22901}
\newcommand*{\VIRGINIAindex}{40}
\affiliation{\VIRGINIA}
\newcommand*{\WM}{College of William and Mary, Williamsburg, Virginia 23187-8795}
\newcommand*{\WMindex}{41}
\affiliation{\WM}
\newcommand*{\YEREVAN}{Yerevan Physics Institute, 375036 Yerevan, Armenia}
\newcommand*{\YEREVANindex}{42}
\affiliation{\YEREVAN}
 \newcommand*{\NOWINFNGE}{INFN, Sezione di Genova, 16146 Genova, Italy}
\newcommand*{\NOWEDINBURGH}{Edinburgh University, Edinburgh EH9 3JZ, United Kingdom}


\author {R.~Nasseripour} \affiliation{\GWU}
\author {B.L.~Berman} \affiliation{\GWU}
\author {K.P. ~Adhikari} 
\affiliation{\ODU}
\author {D.~Adikaram} 
\affiliation{\ODU}
\author {M.~Anghinolfi} 
\affiliation{\INFNGE}
\author {J.~Ball} 
\affiliation{\SACLAY}
\author {M.~Battaglieri} 
\affiliation{\INFNGE}
\author {V.~Batourine} 
\affiliation{\JLAB}
\author {I.~Bedlinskiy} 
\affiliation{\ITEP}
\author {A.S.~Biselli} 
\affiliation{\FU}
\affiliation{\RPI}
% ...

\collaboration{The CLAS Collaboration}
\noaffiliation

%
%
%
%
%
%
%

\title{Coherent Photoproduction of $\pi^+$ from $^3$He}
 
\date{\today}

\begin{abstract}

We have measured the differential cross section for the 
$\gamma$$^3$He$\rightarrow \pi^+ t$ reaction. 
This reaction was studied using the CEBAF Large Acceptance
Spectrometer (CLAS) at Jefferson Lab. Real photons produced
with the Hall-B bremsstrahlung tagging system in the energy range 
from 0.50 to 1.55 GeV were incident on a cryogenic liquid $^3$He
target. The differential cross sections for the 
$\gamma$$^3$He$\rightarrow \pi^+ t$ reaction were measured as a function
of photon-beam energy and pion-scattering angle. 
Theoretical predictions to date cannot explain
the large cross sections except at backward angles, showing that additional
components must be added to the model.
\end{abstract}


\maketitle


\section{Introduction}
\label{sec:intro}

Comparing an elementary meson production process on a free
nucleon with the same process inside a nucleus is 
an interesting problem in nuclear physics.
The contribution of mesonic degrees of freedom to the various 
processes in nuclei can be investigated in the case of the two- 
and three-nucleon systems for which accurate wave functions, 
based on realistic nucleon-nucleon potentials, are available.
Studying this 
production process is ideal for understanding the interaction of pions 
with nuclei and to search for possible effects mediated by nucleon resonances in 
nuclear matter.
Reactions such as 
$\gamma + ^3$He$ \rightarrow \pi^+ + t$, $\gamma + ^3$He$ \rightarrow \pi^0 + ^3$He,
$\gamma + t \rightarrow \pi^- + ^3$He, and $\gamma + t \rightarrow \pi^0 + t$ 
have been studied by both experimental and theoretical groups
over the last four decades(\cite{o'fallon}-\cite{tiator-PRL75-1995}). 
Studying these processes is useful in 
developing our understanding of nuclear structure and the long-range
part of the nucleon-nucleon interaction described by the one-pion exchange 
model. 
However, all the previous measurements were done
near threshold or in the $\Delta$ resonance region.

This measurement is part of a program at Jefferson Lab to study 
the mechanisms of photon-induced reactions in few-body systems. 
This program aims to investigate the fundamental processes in 
the nuclear environment and to test the theoretical calculations 
that are performed using the exact few-body nuclear wave functions 
based on nucleon-nucleon interactions.

The goal of the present analysis is to measure the differential
cross section for the $\gamma + ^3$He $\rightarrow \pi^+ + t$
reaction for energies above the $\Delta$ resonance region. 
This analysis is complementary to the previously reported measurements
on three-body systems, \it{e.g.} \rm the three-body photodisintegration 
of $^3$He \cite{niccolai}.
The  $\gamma + ^3$He $\rightarrow \pi^+ + t$ channel is one of the 
most important pion-production channels because it is an 
isoelastic nuclear transition within the isodoublet ($^3$H,$^3$He)
with the same quantum numbers as the elementary reaction on the 
nucleon. The same nuclear wave functions can be used for the
initial and final states (except for Coulomb effects).
This reaction is particularly attractive because the $^3$He 
target is the lightest nucleus on which one can observe coherent $\pi^+$ 
photoproduction with charge exchange. It allows us to study pion photoproduction 
in a complex nucleus where the final state, consisting of a 
free pion and triton, is well defined and can be identified easily 
in terms of energy and angle or momentum transfer.


The first experiment to measure the cross section for
$\gamma + ^3$He $\rightarrow \pi^+ + t$ over a range of energies and
angles was performed by
O'Fallon {\it et al.} in 1965 \cite{o'fallon}. The measurement was 
done for photon energies of 180 to 260 MeV and triton scattering
angles of 26, 30, 35 and 40 degrees. They found that
the cross section could be described by the cross section from 
a single free proton times the square of the nuclear matter form factor
for $^3$He, modified by kinematic factors. However, the measured 
cross sections were from 25 to 50\% below the simple form-factor theory.
It was suggested that this discrepancy was due to a 
suppression of pion production in nuclear matter. 

In 1979, Argan {\it et al.} \cite{argan} measured the yield of 
$\pi^+$ photoproduction on $^3$He near threshold and compared 
it with electron scattering data on the proton. They obtained
the matrix element for threshold pion photoproduction and showed that 
a unique form factor cannot account for both processes. This 
suggested that many-body contributions affect the two reactions differently.
In fact, to achieve a complete coherent calculation and to obtain 
quantitative information on the many body contribution to pion
photoproduction, it was suggested that the $^3$He and the deuterium 
cases must be treated in parallel. 
Another earlier experiment that measured the differential 
cross section for $\gamma + ^3$He $\rightarrow \pi^+ + t$
was performed by Bachelier {\it et al.} \cite{bachelier} in 1973.
In that experiment, the differential cross section was measured
at a constant value of the momentum transfer of the recoiling
triton using the bremsstrahlung photon beam (227.5 to 453 MeV) of the Saclay
linear electron accelerator. In that work, the experimental results were 
obtained as a function of the incident-photon 
energy and compared with the calculations of Lazard and Maric \cite{lazard}.

Bellinghausen {\it et al.} \cite{bellinghausen} performed an experiment 
in Bonn in 1985 where the photoproduction of charged pions on $^3$He and $^3$H
was measured in the $\Delta$(1232)-resonance region with 
an incident photon energy range of 250 to 450 MeV. The results of that
measurement for $\gamma + ^3$He $\rightarrow \pi^+ + t$
were compared with
the calculation of Sanchez-Gomez and Pascual \cite{sanchez}.
In their model, the photoproduction of pions on nuclei
with three nucleons is considered in the elastic channel.
Calculations were performed using the impulse approximation
and neglecting rescattering effects. These processes were
studied for incident photon energies between 200 and 500 MeV
in the laboratory frame. 

The current analysis is the first to report on the 
$\gamma + ^3$He $\rightarrow \pi^+ + t$ channel with incident photon
energies above 500 MeV. In section \ref{sec:theory} we discuss 
the development of the model calculations. The description of the experiment
and the data analysis procedures including the event selection, background corrections,
study of the detector acceptance, extracting cross sections, 
and the systematic uncertainties are given in section \ref{sec:exper}. 
Section \ref{sec:result} contains the results and comparison with the
model calculations.




\section{Model Predictions}
\label{sec:theory}

On the theoretical front, a model was developed by Tiator {\it et al.}
\cite{tiator} based on realistic three-body Faddeev functions in the
plane-wave impulse approximation (PWIA). 
This model used a production process with Born terms, vector meson 
exchange and $\Delta$(1232) excitation.
Good agreement was found with low-momentum-transfer data (up to 3.1 fm$^{-2}$)
from Ref. \cite{bachelier}, however, the PWIA could not explain the data 
at higher momentum transfer.

In a later calculation performed by Kamalov {\it et al.}
\cite{kamalov}, the intermediate pion scattering between two 
nucleons also has been taken into account. 
In this model, the coherent $\pi^0$ and $\pi^+$ photoproduction and elastic
and charge-exchange pion scattering on $^3$He have been calculated in a consistent 
way. In this model, realistic three-body Faddeev wave functions have been used 
and full non-local distorted-wave impulse-approximation (DWIA) results 
for pion photoproduction were obtained. Comparison with experimental
data showed good agreement over a wide range of momentum transfer for  
the photon energy range between 230 and 450 MeV. 

\begin{figure}[t]
 \begin{center}
 \mbox{\epsfxsize=8.5cm\leavevmode \epsffile{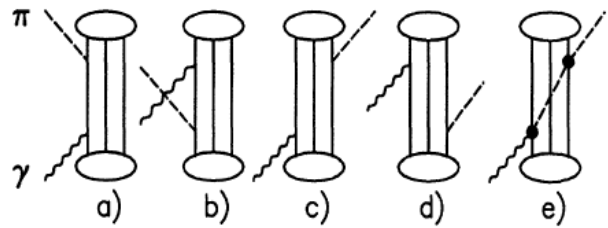}}
 \end{center}
\caption{\small{Diagrams for the dispersive and pion rescattering
terms in nuclear pion photoproduction. The two-body diagrams are shown
in (c)} and (d). Figure is from Ref. \cite{tiator-PRL75-1995}.}
\label{fig:two-body}
\end{figure}

\begin{figure}[htbp]
 \begin{center}
 \mbox{\epsfxsize=7.0cm\leavevmode \epsffile{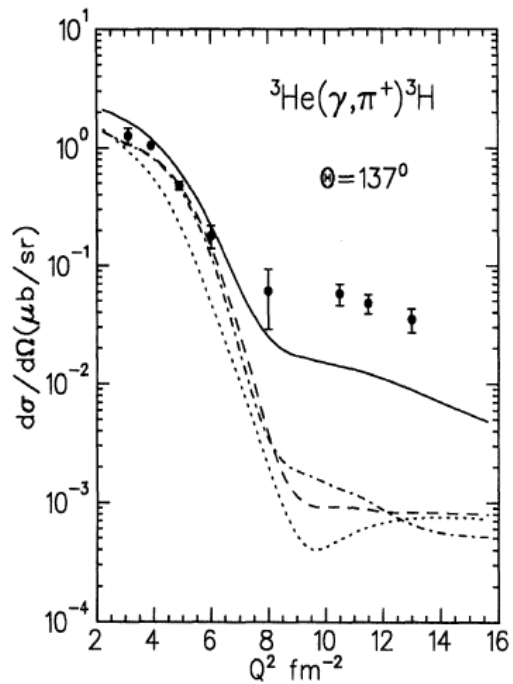}}
 \end{center}
\caption{\small{Differential cross section at
$\theta_{\pi}^{c.m.}=137^\circ$ as a function of nuclear momentum transfer
$Q^2$ from Ref. \cite{bachelier}. The dotted (dashed) curves show the PWIA (DWIA) 
results. The dash-dotted curve includes the corrections due to the coupling
with the breakup channels, and the solid line shows the complete calculation
with the additional two-body mechanisms. Figure is from Ref. \cite{tiator-PRL75-1995}.}}
\label{fig:tiator1995}
\end{figure}

In 1995, the two-body mechanisms were explicitly included in the 
model \cite{tiator-PRL75-1995} where the photon is absorbed by one 
nucleon and the pion is emitted from the other nucleon (Fig. \ref{fig:two-body}). 
The inclusion of these processes resulted in a better agreement between the 
calculations and the previous data at higher momentum transfers.
However, even with all of the considered effects and pion distortions, the model 
could not account for the large enhancement seen in the experimental 
data at large $Q^2$ ($Q^2 >$6 fm$^{-2}$).
Fig. \ref{fig:tiator1995} shows the differential cross section at
$\theta_{\pi}^{c.m.}=137^\circ$ as a function of nuclear momentum transfer
$Q^2$ from Ref. \cite{bachelier}, compared with the complete model calculations
with the additional two-body contribution. The variable $Q^2$ is the 
square of the three-momentum of the recoil triton.
 



\section{Experiment and Data Analysis}
\label{sec:exper}

\subsection{Experimental Apparatus}

The $\gamma ^3$He $\rightarrow t \pi^+$ reaction was measured during CLAS experiment E93-044
(g3a running period) in December 1999
with the CEBAF Large-Acceptance Spectrometer (CLAS) at Jefferson Lab (\cite{mecking}). 
CLAS is a large acceptance spectrometer used to detect multiparticle final states.
Six superconducting coils generate a toroidal magnetic field around the target with
azimuthal symmetry about the beam axis. The coils divide CLAS into six sectors,
each functioning as an independent magnetic spectrometer. Each sector is instrumented
with three regions of drift chambers (DCs), R1-3, to determine charged-particle 
trajectories \cite{dc}, and
scintillator counters (SCs) for time-of-flight measurements \cite{sc}. 
In the forward region, gas-filled threshold Cherenkov counters (CCs) are used for electron/pion
separation up to 2.5 GeV \cite{cc}, and electromagnetic calorimeters (ECs) are used to identify
and measure the energy of electrons and high-energy neutral particles, as well as 
to provide electron/pion separation \cite{ec}.
The primary 1.645 GeV electron beam was incident on the 
thin radiator of the Hall-B Photon Tagger \cite{sober}. 
Tagged photons were produced with 20-95\%
of the energy of the primary electron beam.
In the g3a experiment, real photons tagged in the energy range from 0.35 to 1.55 GeV were incident on 
an 18-cm-thick liquid $^3$He target.
The field of the CLAS toroidal magnet was set to half of its maximum value, 
to optimize the momentum 
resolution and the acceptance for positively charged particles. 
A trigger was used with a required coincidence between hits in the Tagger, 
the Start
Counter (ST), and the time-of-flight (TOF) paddles. About 10$^9$ triggers were collected at the 
production current of 10 nA. 

\subsection{Event Selection}
\begin{figure}[htbp]
 \begin{center}
 \mbox{\epsfxsize=8.0cm\leavevmode \epsffile{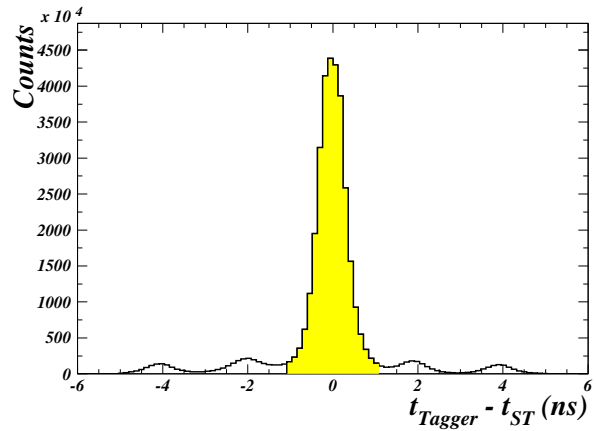}}	
 \end{center}
\caption{\small{Difference between the Tagger time and the Start-Counter (ST) time
(solid histogram). The Tagger and ST coincidence time for selected events 
is required to be within 1 ns (shaded histogram).
The secondary peaks, corresponding to the nearby beam buckets, are also visible.}}
\label{fig:dt}
\end{figure}

In order to associate the reaction of interest with the triggering tagged photon,  
the coincidence time between the Tagger and CLAS was required to 
be within $\pm$1 ns. A cut was applied to the time  
difference, $\Delta t$, between the CLAS start time at the interaction point recorded by
the Start Counter (ST) and the Tagger.  
The central peak in Fig. \ref{fig:dt} corresponds to the tagger hits that are 
in time coincidence with CLAS within the 2-ns-wide 
beam bucket.   
In the g3a run period, only about 2$\%$ of
the events contained more than one tagged photon.

\begin{figure}[t]
 \begin{center}
 \mbox{\epsfxsize=8.0cm\leavevmode \epsffile{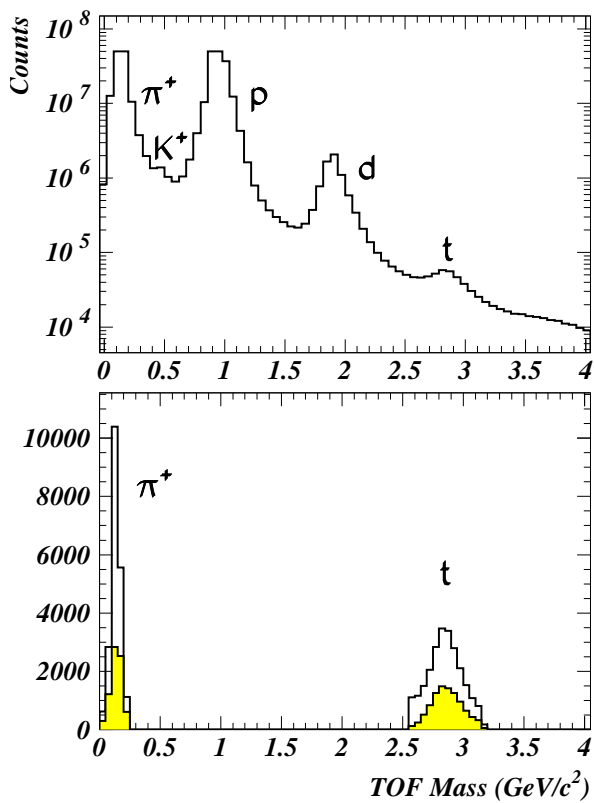}}	
 \end{center}
\caption{\small{Hadron mass calculated from the time-of-flight information.
The histogram in the top panel shows the mass distribution for all positively 
charged hadrons. The solid histogram in the lower panel is the selected sample of pions and 
tritons that are detected in coincidence. The shaded histogram 
shows the same distribution after applying all the kinematical cuts to remove 
the background (see Section \ref{sec:background} for details).}}
\label{fig:tofmass}
\end{figure}

The final-state particles were identified by determining their charge, momentum,
and velocity. Charge and momentum were obtained from the drift-chamber 
tracking information and the velocity from the time of flight
and path length to the scintillation counters. Figure \ref{fig:tofmass}
shows the reconstructed mass distribution of positively charged particles. The 
events of interest were those with two and only two positively charged 
particles detected in coincidence. A triton candidate was required to
have a positive charge and a reconstructed mass squared $m^2$ between 6.5 
and 10.0 (GeV/{\it{c}}$^2$)$^2$. A pion candidate
was required to have a positive charge and a reconstructed mass
squared between 0.05 and 0.3 (GeV/{\it{c}}$^2$)$^2$.
In order to assure that the events of interest were produced within the
$^3$He target volume, a cut was applied to the $z$-component of the interaction 
vertex along the beam line.

\begin{figure}[htbp]
 \begin{center}
 \mbox{\epsfxsize=9.0cm\leavevmode \epsffile{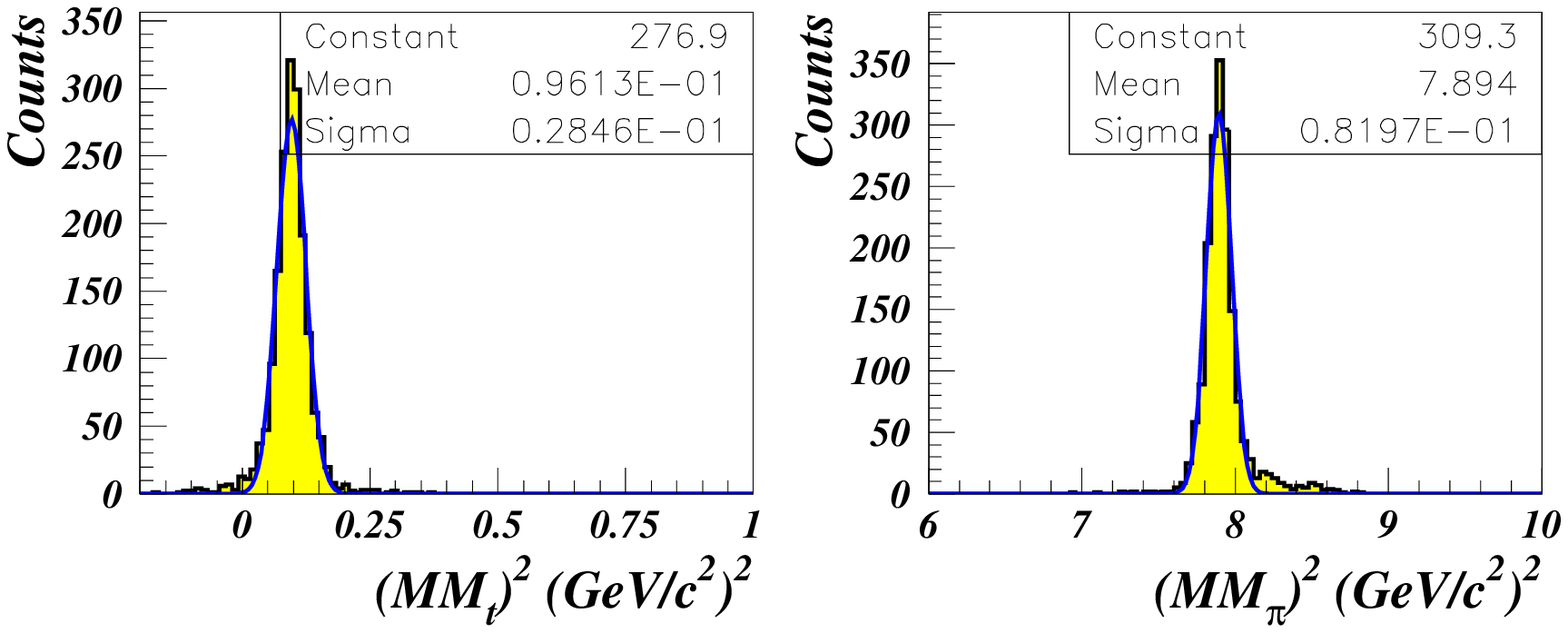}}
 \mbox{\epsfxsize=9.0cm\leavevmode \epsffile{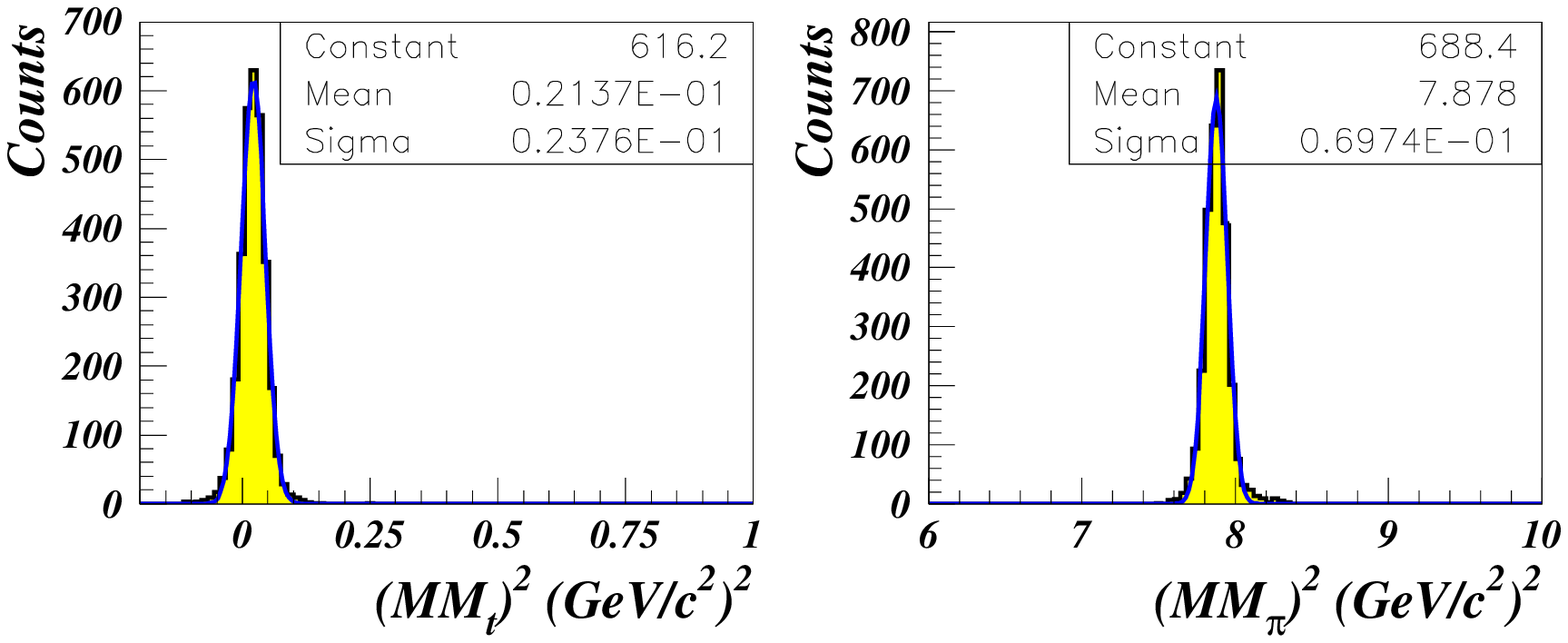}}		
 \end{center}
\caption{\small{Distributions of the missing mass squared of the detected triton (left) and pion (right) 
before (top) and after (bottom) the energy-loss corrections. Gaussian fits shown in solid blue 
are performed to determine the mean value and the width of each distribution. The accepted
values for the pion and triton mass squared are 0.0195 and 7.890 (GeV/{\it{c}}$^2$)$^2$, respectively.}}
\label{fig:eloss}
\end{figure}
Energy-loss corrections were applied to the selected particles because
they lose a non-negligible part of their energy in the target material
and start counter before they reach the drift chambers.  
The effect of the energy-loss corrections after
applying all of the kinematic cuts on the final sample of $t\pi^+$ data 
is shown in Fig. \ref{fig:eloss}.  The importance  
of these corrections can be demonstrated by comparing the missing-mass squared of 
either the detected pion or the detected triton before and after applying 
these corrections. 


Also, 
fiducial-volume cuts were applied to ensure that the particles are detected
within those parts of the volume of CLAS where the detection efficiency is 
high and uniform.
These cuts select regions of CLAS where simulations reproduce the
detector response reasonably well.

\subsection{Background Corrections}
\label{sec:background}
In order to select cleanly the $\gamma ^3$He $\rightarrow t\pi^+$ channel,
two-body kinematics were used. The two-body final-state kinematics 
for real events require that the missing energy, missing momentum,
and missing-mass squared for $t\pi^+$ events be zero.  
Also, the opening angle between the three-vectors of the detected pion and triton
$\theta_{t\pi^+}$ should be close to 180$^{\circ}$ in the center-of-mass frame.
Our initial sample of events contains two and only two charged particles.
Four-vector conservation for the reaction $\gamma ^3$He $\rightarrow t\pi^+$,
leads to the determination of three 
kinematic variables --  
the missing energy $E_X$, the missing momentum
$P_X $, 
and the missing-mass squared $M_X^2 = E_X^2 - P_X^2$. 
These kinematic variables are plotted in 
Fig. \ref{fig:cuts1}. For the real coherent $t\pi^+$ events, we then have 
$E_X=0$ GeV, $P_X= 0$ GeV/{\it{c}}, $M_X^2 = 0$ (GeV/{\it{c}}$^2$)$^2$, and $\theta_{t\pi^+}= 180^{\circ}$.
Indeed, in Fig. \ref{fig:cuts1} one can see clear peaks showing the real coherent 
$t\pi^+$ events.
However, some background can be seen in the selected events.
These events (mostly due to the $t\pi^+\pi^0$ channel) can be removed by applying additional 
kinematic cuts as follows: 

\begin{figure}[t]
 \begin{center}
 \mbox{\epsfxsize=9.0cm\leavevmode \epsffile{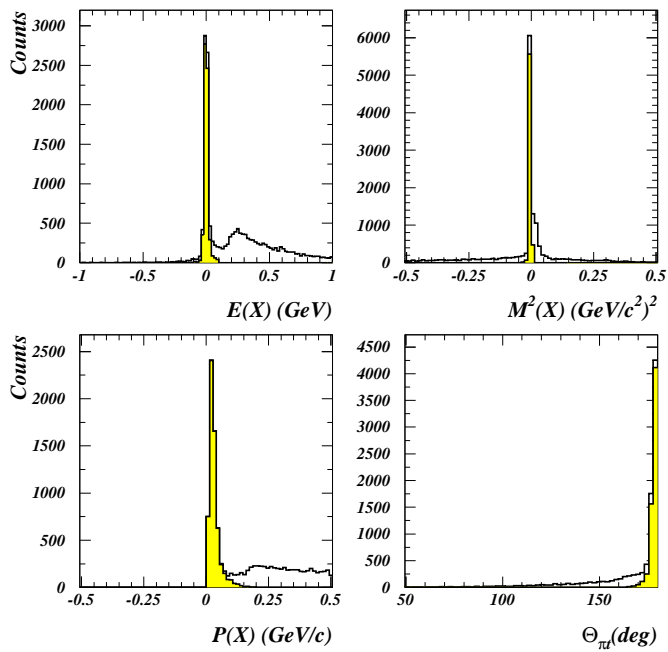}}	
 \end{center}
\caption{\small{The $\pi^+ t$ two-body final-state kinematics require the missing energy
(upper left), missing mass squared (upper right), and missing momentum (lower left)
to be zero, and the $\pi^+ t$ opening angle (lower right) to be 180$^o$. The peaks correspond
to the real coherent $\pi^+ t$ events from the $^3$He target.
The shaded areas correspond to the nearly background-free sample of $\pi^+ t$ events after 
the five kinematical cuts described in the text were applied.}}
\label{fig:cuts1}
\end{figure}
\begin{figure}[t]
 \begin{center}
 \mbox{\epsfxsize=9.0cm\leavevmode \epsffile{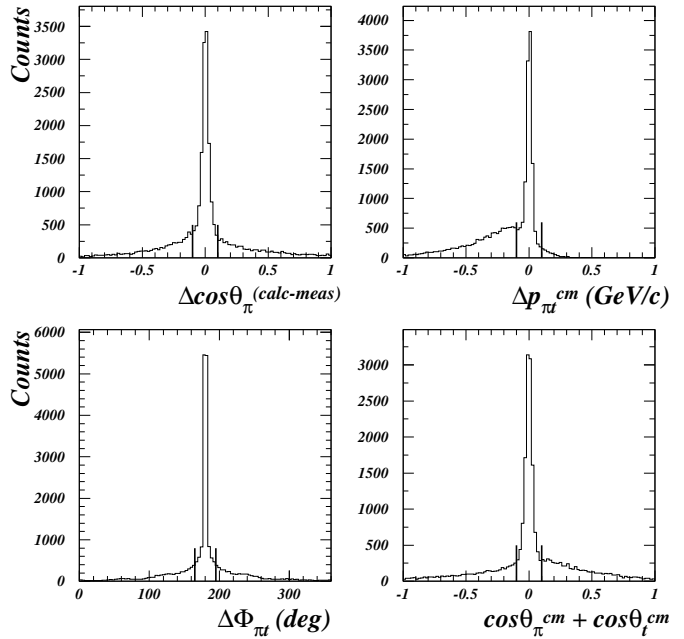}}	
 \end{center}
\caption{\small{Cuts were applied on various kinematical variables to remove
the background. Upper left: The difference between the measured and calculated
pion scattering angles. Upper right: The difference
between the magnitude of pion and triton momenta. Lower left: The difference between
the pion and triton azimuthal angles. Lower right: The sum of the cosines of the pion and 
triton scattering angles. All quantities are shown in the center-of-mass frame.}}
\label{fig:cuts2}
\end{figure}


\begin{enumerate}
\item The first cut is applied to the difference between the measured scattering
angle of the pion in the center-of-mass frame (from the measured three-momentum 
vector of the pion) and the calculated one from the conservation of the four-momenta 
in the $\gamma ^3$He $\rightarrow t\pi^+$  reaction (by measuring only the 
triton momentum). This difference is plotted in the upper-left side of 
Fig. \ref{fig:cuts2}. The clear peak around zero
corresponds to the real events from the coherent production of a pion and a triton. 
The events for which this angular difference is outside of the range [-0.1,0.1] 
were removed from the data.

\item The second cut is applied to the difference between the momenta of the pion 
and the triton in the center-of-mass frame. For the real $t\pi^+$ events, this difference 
shows a peak around zero with a tail that could be due to the $t\pi^+\pi^0$ events,
as shown in the upper-right panel of Fig. \ref{fig:cuts2}.
The applied cut requires this difference to be between -0.1 and 0.1 GeV/c.

\item The third cut requires the pion and triton three-momenta
to be in the same plane--as the initial photon--, \it{i.e.}, \rm the difference between the azimuthal angles for 
the pion and the triton in the center-of-mass frame is selected to be 
165$^{\circ}<\phi_{t\pi^+}^{cm}<$195$^{\circ}$. This distribution
is shown in the lower-left panel of Fig. \ref{fig:cuts2}. A prominent peak around 180$^{\circ}$
is clearly seen.

\item The fourth cut is applied to the sum of the cosines of the pion and triton
scattering angles in the center-of-mass frame, shown in the lower-right panel of Fig. \ref{fig:cuts2}. 
This cut retains only those events with -0.1$<$cos$\theta_\pi^{cm}+$cos$\theta_t^{cm}<$0.1.

\item Finally, the fifth cut requires the $t\pi^+$ missing energy to be 
-0.1$<E(X)<$0.1 GeV, shown in the upper left panel of Fig. \ref{fig:cuts1}. 

\end{enumerate}

\begin{figure}[t]
 \begin{center}
 \mbox{\epsfxsize=6.5cm\leavevmode \epsffile{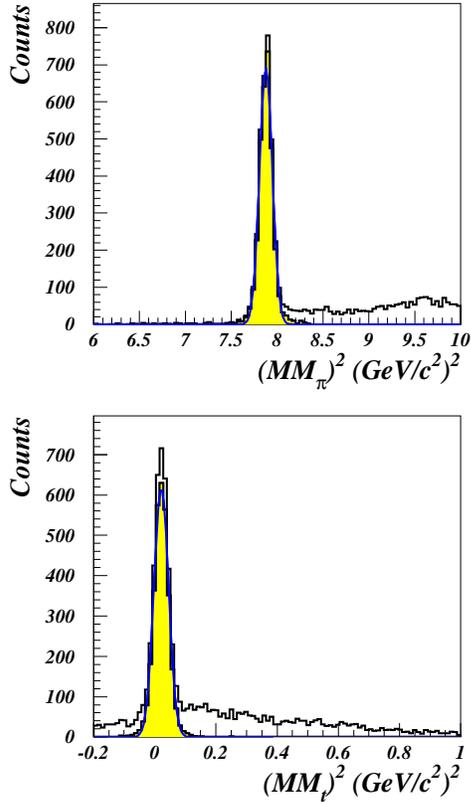}}	
 \end{center}
\caption{\small{The calculated values for the missing mass squared for 
the detected pion (top) and the detected triton (bottom), before (solid histogram) 
and after (shaded histogram) applying the kinematical cuts. 
The background is removed by the kinematical cuts.}}
\label{fig:mm}
\end{figure}

The value of each of these cuts is optimized such that the maximum number of ``good''
$t\pi^+$ events is retained. 
Using these cuts, the background in the spectra of the previously
described kinematic variables is mostly removed, as can be seen for 
the shaded areas of Fig. \ref{fig:cuts1}.
The sample of events used after these cuts is nearly background-free.
This is further supported by calculating the missing-mass squared of 
either the detected 
pion or the detected triton. These distributions are shown before and 
after the above cuts in Fig. \ref{fig:mm}, and show that the background has been  
removed. The clean sample of pions and tritons that are detected in coincidence is also
shown within the shaded areas of Fig. \ref{fig:tofmass}. 

Table \ref{tab:cuts} summarizes the final cuts used to identify the
$t\pi^+$ events as described in this section.
\begin{figure}[t]
 \begin{center}
 \mbox{\epsfxsize=7.5cm\leavevmode \epsffile{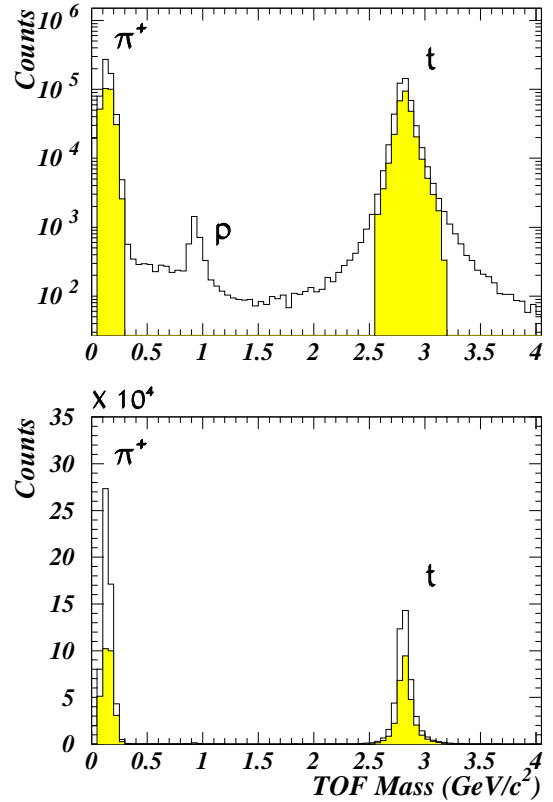}}
 \end{center}
\caption{\small{Simulated TOF masses for Monte-Carlo generated
events, plotted for both logarithmic (top) and linear (bottom) scales, before (solid histogram) 
and after (shaded areas) applying all of the cuts.}}
\label{fig:tofmass-sim}
\end{figure}
\begin{table}[h]
\begin{center}
\caption[]{\small{Summary of kinematic cuts for event selection.}}
\label{tab:cuts}
\begin{tabular} {lc} \hline \hline
Description & Cut  \\ \hline
Coincidence time $\Delta t$	& $<$ 1 nsec   \\ 
Positively charged particles	& 2 			  \\ 
Pion identification		& $-0.06<m_\pi^2<0.05$ (GeV/{\it{c}}$^2)^2$ \\ 
Triton identification   	& $6.5<m_t^2<10.0$ (GeV/{\it{c}}$^2)^2$ \\ 
z-vertex 			& [-8,7.5] (cm)		\\ 
$\Delta \cos\theta_\pi^{cm}$	& [-0.1,0.1]		\\ 
$\Delta p_{\pi,t}^{cm}$		& [-0.1,0.1] (GeV/{\it{c}})	\\ 
$\Delta \phi_{\pi,t}^{cm}$	& [165,195] deg	\\ 
$\cos\theta_\pi^{cm}+\cos\theta_t^{cm}$	& [-0.1,0.1]	\\ \hline \hline
\end{tabular}
\end{center}
\end{table}
\subsection{Detector Efficiency and Acceptance}

The raw $t\pi^+$ yields are obtained as a function of the photon beam
energy $E_{\gamma}$ and the pion polar angle in the center-of-mass frame $\theta_\pi^{cm}$. 
The yields are corrected for the
detector acceptance using a Monte-Carlo simulation of phase-space-distributed
$t\pi^+$ events within the entire 4$\pi$ solid angle. The photon energy
was generated randomly with a uniform distribution from 0.35 to 1.55 GeV.
The standard GEANT-based CLAS simulation package \cite{gsim} 
was used to simulate the detector response. 
The simulated events were processed with the same event-reconstruction
software that was used to reconstruct the real data. 
Figure \ref{fig:tofmass-sim} shows the reconstructed mass distributions for the 
simulated events with one pion and one triton after applying all of the cuts.

The acceptance is defined as the 
ratio of the number of reconstructed events to the number of generated
events.
Owing to the geometry and the structure of CLAS, there are regions of
solid angle that are not covered by the detector.
Furthermore, the inefficiencies in the various components of the detector
affect its acceptance and, consequently, the event reconstruction in CLAS. 
The acceptance correction factors are obtained as functions of pion
angle $\theta_\pi^{cm}$ and photon energy $E_{\gamma}$
for each kinematic bin 
and are used to 
convert the raw yields into unnormalized cross sections. 



\subsection{Cross Sections}
The differential cross section is obtained from the expression

\begin{equation}
{{d\sigma \over d\Omega} = { N \over \eta_a N_{\gamma} N_T \Delta\Omega}}\;, 
\end{equation}
where $N$ is the number of measured events in a given energy and angular bin of 
solid angle $\Delta\Omega = 2\pi\Delta \cos\theta_{cm}$. 
The CLAS acceptance is given by $\eta_a$; $N_{\gamma}$ is the number of 
photons within the given energy range incident on the target; and $N_T$
is the number of target nuclei per unit area. 

The number of target nuclei per unit area $N_T$ is determined from

\begin{equation}
N_T = {\rho l N_A \over A} \approx 2.089 \times 10^{-10} \;\; \rm{nb}^{-1},
\end{equation}
where $l = 155.0$ mm is the target length, $\rho = 0.0675$ g/cm$^3$ is 
the density of liquid $^3$He, $A = 3.016$ g/mole is its atomic
weight, and $N_A = 6.022 \times 10^{23}$ atoms/mole is Avogadro's
number. 

The photon yield $N_{\gamma}$ was obtained from the tagger hits using the gflux 
analysis package
\cite{ball}. This number is corrected for the data-acquisition dead time. 


\unboldmath

\boldmath
\subsection{Systematic Uncertainties}
\label{sec:syserr}
\unboldmath
Table \ref{tab:sys} summarizes the systematic uncertainties. The uncertainty in the 
photon-flux determination, including the tagger-efficiency evaluation,
is the same as in the g3a analysis of Niccolai {\it et al.} \cite{niccolai}.
The value of the target density given in the literature was used; its uncertainty 
is no larger than 2$\%$. 
The uncertainties due to the 
fiducial cuts are estimated and have been found to be 
negligible.
\begin{table}[h]
\caption[]{\small{Summary of systematic uncertainties arising from various sources.}}
\label{tab:sys}
\begin{tabular} {lc} \hline \hline
Source			& Uncertainty ($\%$)    \\ \hline 
Photon flux		& 6 			\\ 
Target density		& $<$ 2 			\\ 
Fiducial cuts		& negligible		\\ 
Solid angle 		& negligible 		\\ 
CLAS acceptance		& $<$ 15			\\ 
Kinematic cuts 		& $<$ 10                   \\
Timing cut		& negligible			\\    
\hline
Total			& $<$ 20			\\ \hline \hline			
\end{tabular}
\end{table}
The systematic uncertainty due to the CLAS acceptance was obtained
by comparing the cross sections measured by each pair of the
CLAS sectors independently ({\it i.e.}, the data from sectors 1 and 4, 2 and 5,
and 3 and 6 were combined). The mean deviation between the three
sets of cross sections is given in Table \ref{tab:sys}.

In order to estimate the systematic uncertainty due to applying the
kinematic cuts, two sets of altered cuts, loose and tight, were used
and compared with the nominal cuts. The root mean square of
the distribution of the differences between the cross sections obtained
with loose, tight, and the nominal cuts is considered to be a measure of the systematic
uncertainty due to these cuts.

The CLAS acceptance and kinematic cuts constitute the 
largest part of the systematic uncertainty.  
The individual systematic uncertainties are summed in quadrature to less 
than 20$\%$. The statistical uncertainties for the results of many kinematic bins
are larger than
the systematic uncertainties.


\section{Results and Discussion}
\label{sec:result}

\subsection{Cross Sections}
\begin{figure*}[htbp]
 \begin{center}
\mbox{\epsfxsize=13.9cm\leavevmode \epsffile{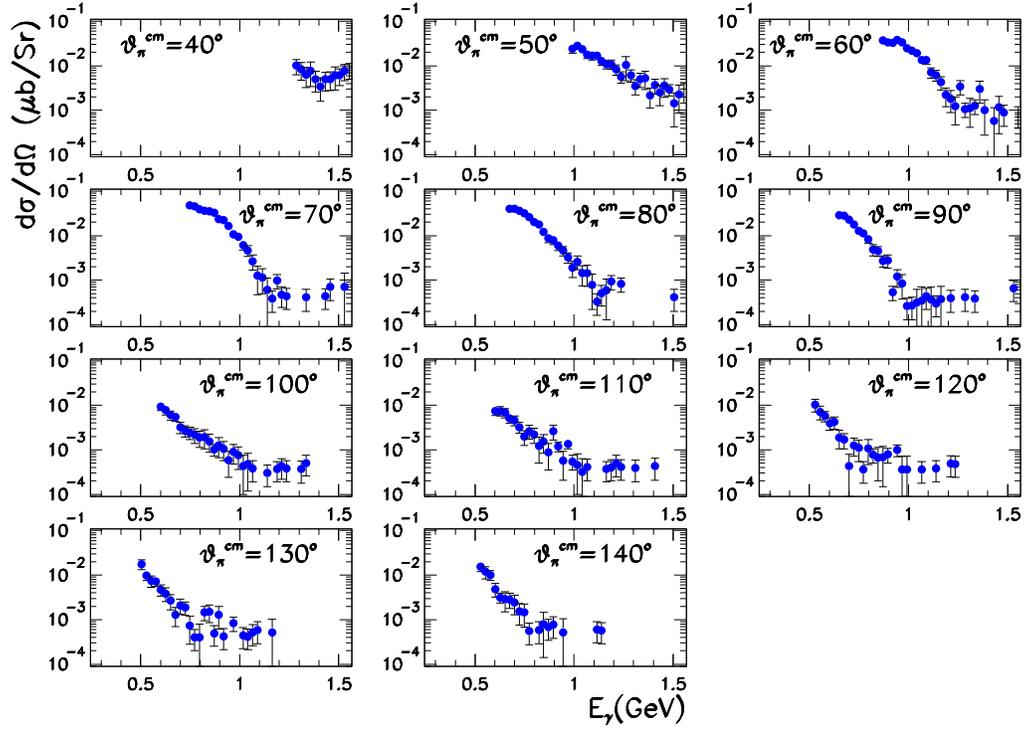}}
 \end{center}
\caption{\small{(Color online) Measured differential cross sections as a function of $E_\gamma$ for
$\theta_\pi^{cm}$= 40, 50, 60, 70, 80, 90, 100, 110, 120, 130, and 140 degrees.
The error bars indicate statistical uncertainties only.}}
\label{fig:xsec-energy}
\end{figure*}
\begin{figure*}[htbp]
 \begin{center}
\mbox{\epsfxsize=13.9cm\leavevmode \epsffile{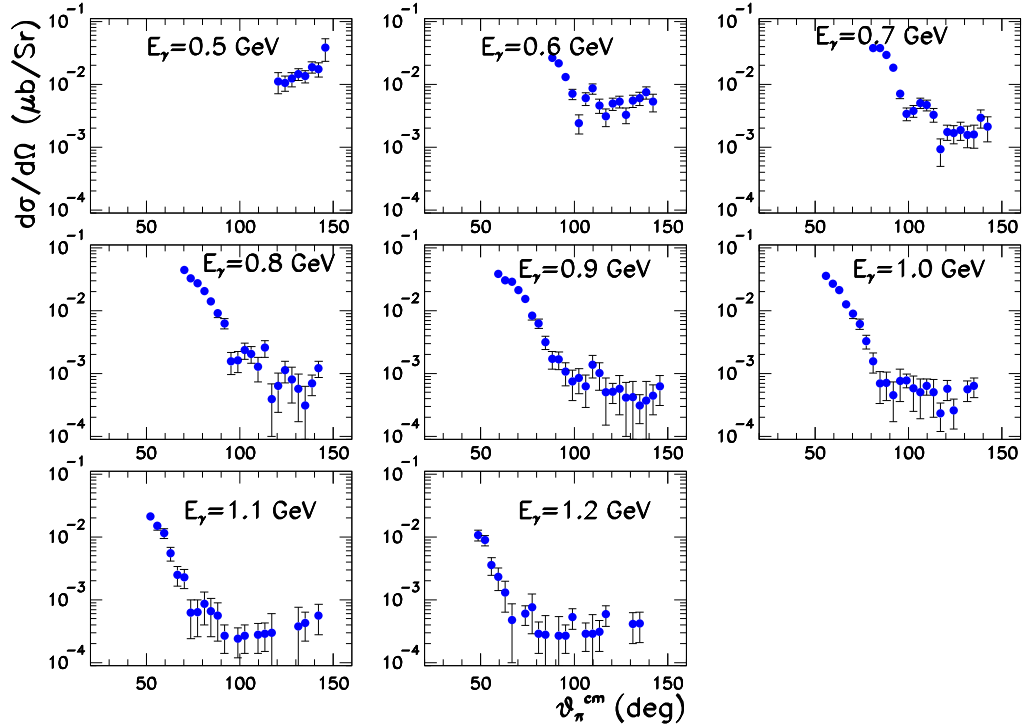}}
 \end{center}
\caption{\small{(Color online) Measured differential cross sections as a function of 
$\theta_\pi^{cm}$ for $E_\gamma$ = 0.5, 0.6, 0.7, 0.8, 0.9, 1.0, 1.1, and 1.2 GeV.
CLAS acceptance limits the detection of the small angle poins.
The error bars indicate statistical uncertainties only. }}
\label{fig:xsec-angle}
\end{figure*}
The measured differential cross sections are shown in Figs. \ref{fig:xsec-energy} 
and \ref{fig:xsec-angle} as functions of photon energy and pion angle,
respectively.
These plots show that the peak of the angular
distributions shifts towards smaller angles with increasing 
photon energy. We have also studied the dependence of the cross
sections on the momentum transferred to the triton, $Q^2$.
This variable 
enters the nuclear wave functions and is mostly responsible for nuclear
structure effects. Our measurements covers a range of $Q^2$=10-37 fm$^{-2}$ 
(0.4-1.5 (GeV/c)$^2$) (see below).


\subsection{Comparison with Model Calculations and Previous Data}

\begin{figure}[t]
 \begin{center}
\epsfig{figure=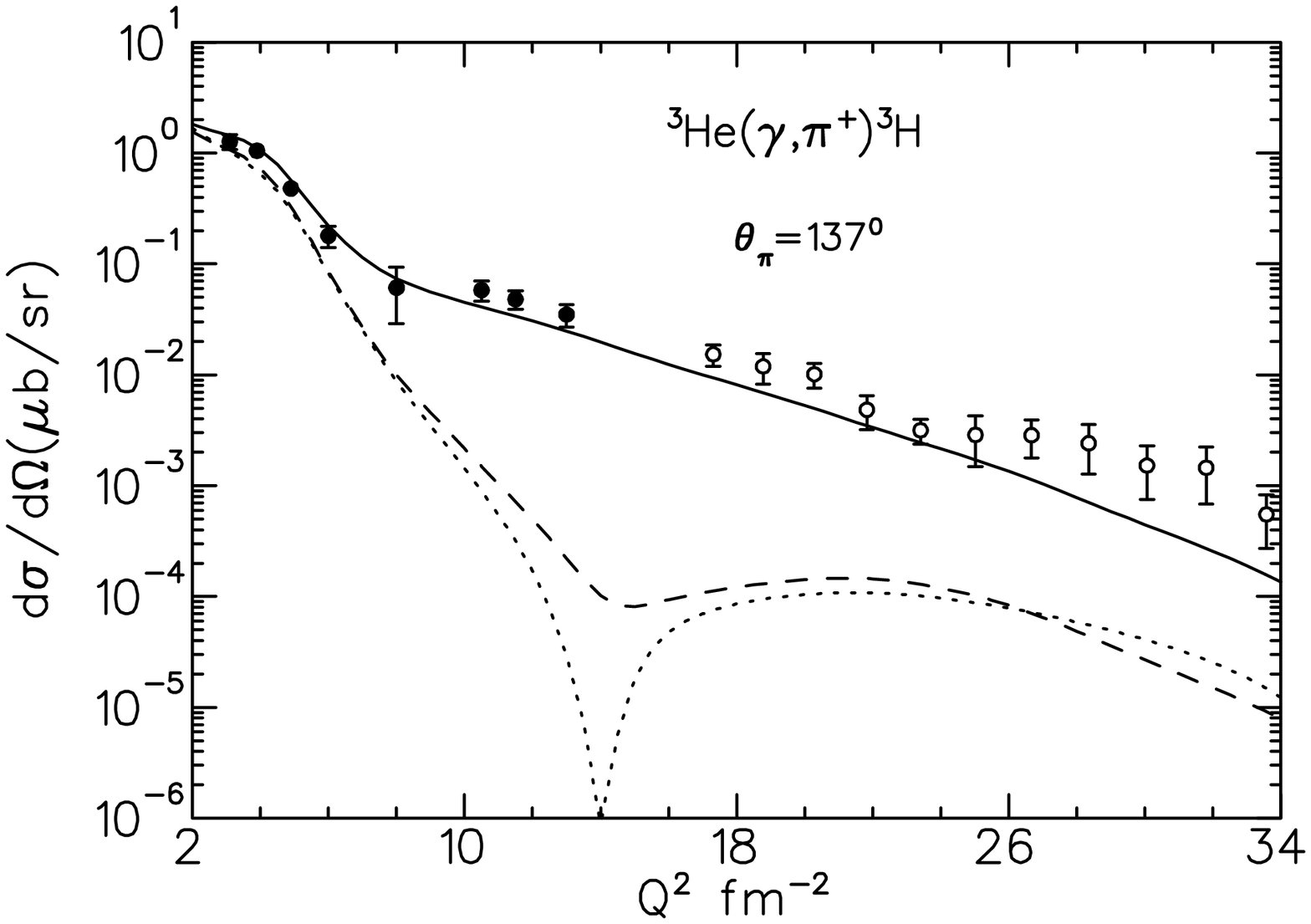, angle=90, width=8.0cm}
 \end{center}
\caption{\small{Momentum-transfer dependence of the differential cross section
for a fixed pion angle of 137 degrees in the c.m. frame. 
The curves show the calculations by Tiator and Kamalov for three different
assumptions: plane-wave impulse approximation PWIA (dotted lines);
distorted-wave impulse approximation DWIA (dashed lines); and 
DWIA + 2-body mechanism \cite{tiator-PRL75-1995} (solid lines).
Our data from CLAS are shown as
open circles
and from Ref. \cite{bachelier} as filled circles.}}
\label{fig:137deg}
\end{figure}

\begin{figure}[t]
 \begin{center}
\epsfig{figure=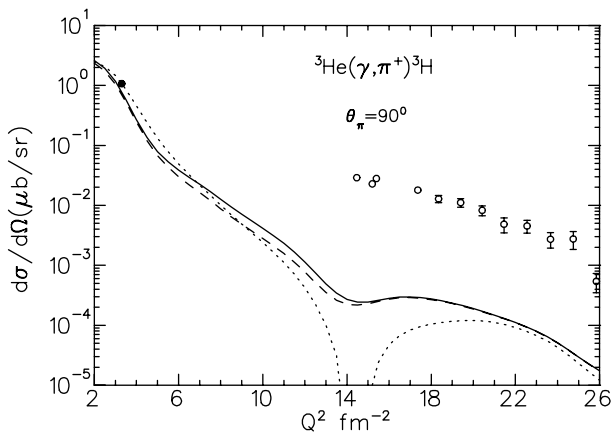, angle=90, width=8.0cm}
 \end{center}
\caption{\small{Momentum-transfer dependence of the differential cross section
for a fixed pion angle of 90 degrees in the c.m. frame. 
The curves are described in Fig. \ref{fig:137deg}. Our data from CLAS are shown as open circles
and from Ref. \cite{d'hose} as a filled circle.}}
\label{fig:90deg}
\end{figure}

\begin{figure}[t]
 \begin{center}
\epsfig{figure=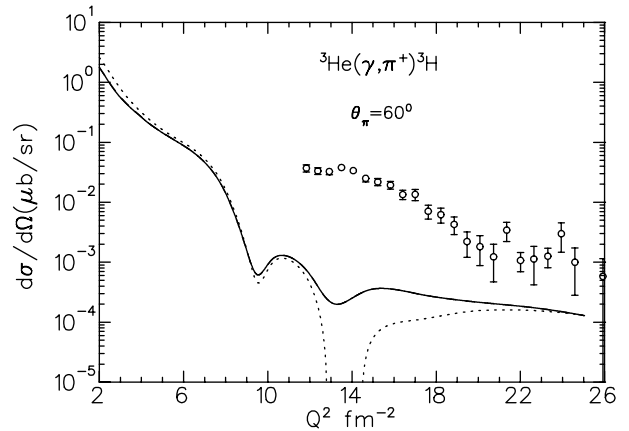, angle=90, width=8.0cm}
 \end{center}
\caption{\small{Momentum-transfer dependence of the differential cross section
for a fixed pion angle of 60 degrees in the c.m. frame. 
The curves are described in Fig. \ref{fig:137deg}. Our data from CLAS are shown as open circles. 
Note that in the forward direction, 
the DWIA and the DWI+2body calculations coincide as expected, so
the 2-body mechanism included in the model does not contribute.}}
\label{fig:60deg}
\end{figure}

\begin{figure*}[htbp]
 \begin{center}
\epsfig{figure=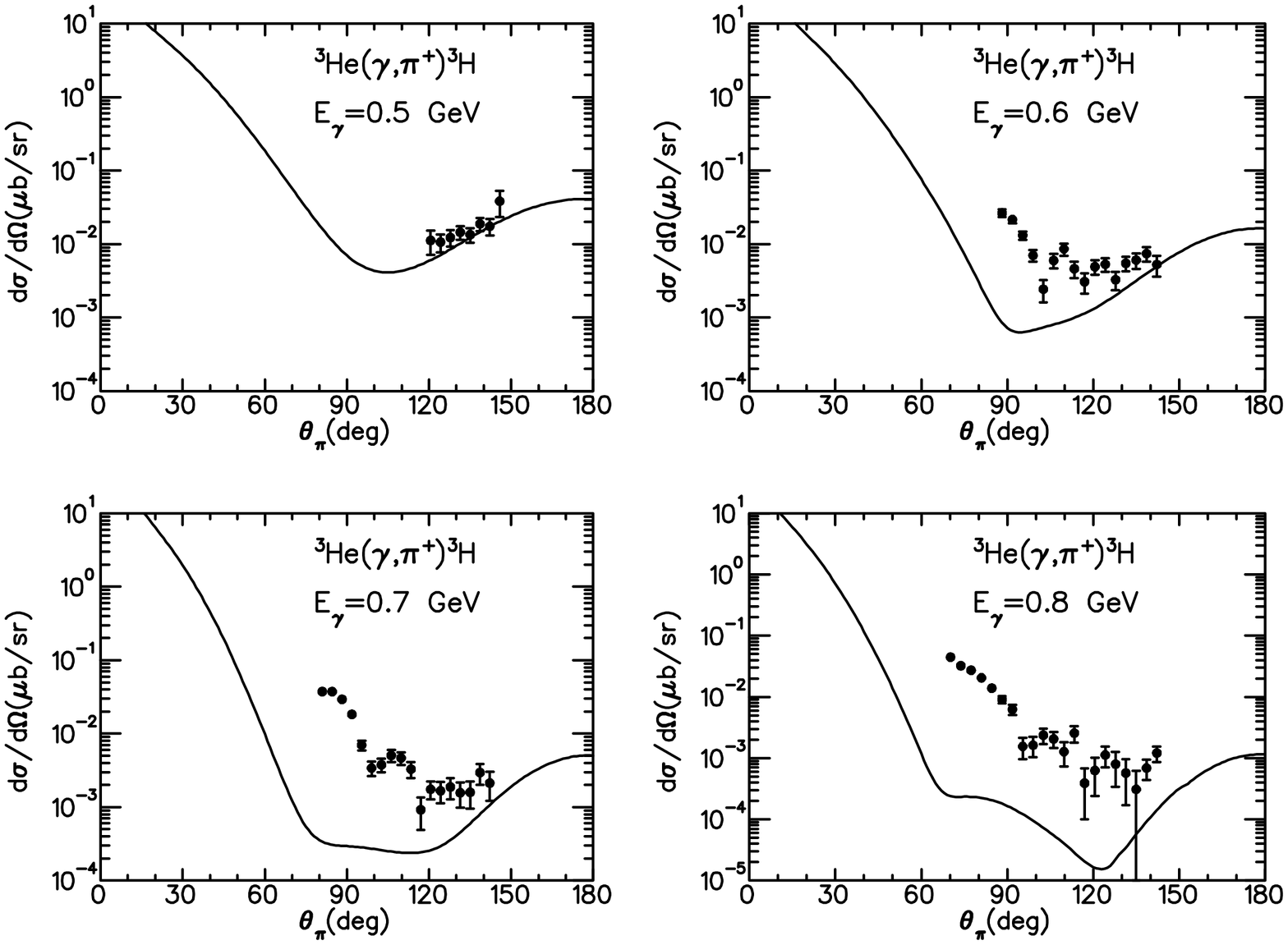, angle=90, width=12.5cm}
 \end{center}
\caption{\small{Comparison of the model calculations from Tiator and 
Kamalov with the differential cross section as a function of pion
scattering angle in the c.m. frame for various photon-energy bins. The model includes 
DWIA + 2-body mechanism (see text).}}
\label{fig:fig32}
\end{figure*}

In this section our results are compared with the model calculations by
Tiator and Kamalov and with previous measurements. 
The calculations were originally suited only 
for the energies from threshold to the $\Delta$ resonance region. Recently this 
model has been extended (with MAID) to higher energies \cite{tiator-private} 
(see Figs. \ref{fig:137deg}-\ref{fig:fig32}). The curves show plane-wave 
impulse approximation PWIA (dotted lines),
distorted-wave impulse approximation DWIA (dashed lines), and 
the DWIA + 2-body mechanism \cite{tiator-PRL75-1995}(solid lines).

There is good agreement between the calculations and experimental data 
for small momentum transfers.
For larger momentum transfers the calculations can describe 
the data only at backward angles. The old measurement at 137 degrees can be 
nicely extended with our data up to $Q^2=$34 fm$^{-2}$ or 1.4 GeV$^2$ (Fig.
\ref{fig:137deg}).
For other angles a huge discrepancy is found, {\it e.g.}, at 90 or 60
degrees (Figs.\ref{fig:90deg} and \ref{fig:60deg}).
With the new elementary production operator from MAID the agreement
with data from Bachelier \it{et al.} \rm \cite{bachelier} is much
improved compared to the previous calculations in 1995 (see Fig. \ref{fig:tiator1995}).

These are interesting results, which were not observed before when
only high-$Q^2$ data were available at one angle, namely 137 degrees. 
Our new data suggest that there are other mechanisms that 
produce much larger contributions than the
1-body (impulse approximation) and the 2-body mechanisms that were proposed in Ref.
\cite{tiator-PRL75-1995}.
It is possible that two- or even three-body effects are
driving the large cross sections, but it is not precisely known 
to what extent.

Figure \ref{fig:fig32} shows the comparison of the angular dependence of our 
cross sections with the full model calculations for four bins of photon energy
from 0.5 to 0.8 GeV. In general, the calculations fail to describe 
our data at higher photon energies and forward angles.
This suggests that the one- and two-body mechanisms alone cannot describe 
our data and that the discrepancy between the data and the calculation might 
be most likely due to the fact that the three-body mechanisms are not included 
in the model.
In fact, strong evidence from analyzing CLAS data in other channels, for example,
$\gamma ^3$He $\rightarrow ppn$ \cite{niccolai}, $\gamma ^3$He $\rightarrow pd$ \cite{yordanka}, 
and $\gamma ^4$He $\rightarrow pt$ \cite{rakhshapt}, suggests that 3-body 
contributions become more important, especially at $E_{\gamma}$=0.6-0.8 GeV.

The models could be improved by including 2-body and 3-body 
meson-exchange currents (MEC). These processes become more important especially
at high momentum transfers because the momentum is shared between two or three
nucleons. The very first attempt to include a two-body MEC in the pion-photoproduction
model, where a pion is exchanged between the two nucleons, was considered by 
Raskin {\it et al.} \cite{raskin}. In that model, a formalism for the pion-photoproduction
amplitude with binding-induced contributions was given. However, no numerical
calculation was ever performed.

Drechsel {\it et al.} \cite{drechsel} and Strueve {\it et al.} \cite{strueve}
also considered the two- and three-body MEC in their calculations for 
the $^3$He and $^3$H form factors. Both models described the experimental
data with a good degree of success after including these processes.
    
Another possible process to include in the model would be the photo-induced
reaction $\Delta(\gamma,\pi N)$ on a free $\Delta$ that is created from the
$N+N \rightarrow \Delta+N$ reaction. The existence of these pre-formed $\Delta$s was
investigated by studying reactions such as $A(\gamma,\pi^+p)B$. It was shown
that the assumption of a small amount of pre-formed $\Delta$ can fit  $^{12}C(\gamma,\pi^+p)$ 
data from MAMI if the $\Delta^{++}$ is in an $S_{3 \over 2}$ orbital \cite{chang}.
Pre-formed $\Delta$s were also introduced in the calculations
of the $^3$He and $^3$H form factors \cite{strueve}.

On the experimental side, it would be interesting to see whether there is a similar 
enhancement in the coherent $\pi^0$ photoproduction cross section at high momentum transfer 
from deuterium \cite{yordankapi0}, $^3$He, and $^4$He targets. Perhaps data are 
available to be analyzed for this channel from various experimental groups, for 
example Crystal Ball in Mainz, Crystal Barrel in Bonn, and CLAS at Jefferson Lab.


In summary, we have measured the differential cross section for 
the $\gamma ^3$He $\rightarrow t\pi^+$ reaction
in the energy range from 0.5 to 1.55 GeV, for pion center-of-mass 
angles between 40 and 140 degrees. 
We have compared our data with the results of the only 
available theoretical calculations for these energies 
\cite{kamalov,tiator-PRL75-1995,tiator-private}. 
The comparison shows that the calculations cannot describe our data
at large momentum transfer and measured forward angles. This strongly suggests
that there are additional production mechanisms that are not included 
in the current formulation of the model.  It would certainly be 
interesting to see whether the coherent $\pi^0$ photoproduction shows 
similar effects.


\section*{Acknowledgments}

We would like to acknowledge the outstanding efforts of the staff of 
the Accelerator and the Physics Divisions at Jefferson Lab that made 
this experiment possible.  This work was supported by the U.S.
Department of Energy under grant DE-FG02-95ER40901.
the National Science Foundation, the Italian 
Istituto Nazionale di Fisica Nucleare, the French Centre National de la 
Recherche Scientifique, the French Commissariat \`{a} l'Energie 
Atomique, the National Reseach Foundation of Korea,   
the UK Science and Technology Facilities Council (STFC), and 
Scottish Universities Physics Alliance (SUPA).
The Southeastern Universities Research Association (SURA) operated the 
Thomas Jefferson National Accelerator Facility for the United States 
Department of Energy under contract DE-AC05-84ER40150. 


\end{document}